\documentclass[draft,grl]{agutex}

\usepackage[dvips]{graphicx}

\setkeys{Gin}{draft=false}

\authorrunninghead{ESELEVICH AND ESELEVICH}

\titlerunninghead{A POSSIBLE MECHANISM OF ENERGY DISSIPATION}

\authoraddr{M. V. Eselevich,
Institute of Solar-Terrestrial physics, Lermontova str. 126a,
P.O.Box 291, Irkutsk, 664033, Russia. (mesel@iszf.irk.ru)}

\authoraddr{V. G. Eselevich,
Institute of Solar-Terrestrial physics, Lermontova str. 126a,
P.O.Box 291, Irkutsk, 664033, Russia. (esel@iszf.irk.ru)}

\begin{document}

\title{A possible mechanism of energy dissipation
in the front of a shock wave driven ahead of a coronal mass ejection}

\authors{M. V. Eselevich \altaffilmark{1}
and V. G. Eselevich \altaffilmark{1}}

\altaffiltext{1}{Institute of Solar-Terrestrial physics, Irkutsk,
Russia}

\begin{abstract}

Mark 4 and LASCO C2, C3 coronagraph data analysis shows that, up
to the distance $R\sim$ 5 R$_\odot$ from the center of the Sun,
the thickness of a CME-generated shock front may be of order of
the proton mean free path. This means that the energy dissipation
mechanism in a shock front at these distances is collisional.
\end{abstract}

\begin{article}

\section{Introduction}

Eselevich M. and V. \citep{Eselevich2008} revealed that there is a
disturbed region extended along the direction of coronal mass
ejection (CME) propagation in front of the CME, when its velocity
$u$ relative to the ambient coronal plasma is below a certain
critical velocity $u_C$. Given $u > u_C$, a discontinuity in the
difference brightness
distributions is formed in the frontal part of the disturbed
region. Since $u_C$ is close to the local fast-mode MHD velocity
the formation of such
a discontinuity may be associated with shock wave formation.
If we managed to resolve this discontinuity in space, determining
its thickness $\delta _F$ (by measuring the shock front profile,
in essence), we could be able to clarify the dissipation mechanism
in the shock front in the corona.

The purpose of this paper is: 1) to justify the correctness of
$\delta _F$ measurements in the solar corona using Mark 4 and
LASCO C2; 
2) to discuss a possible dissipation mechanism in
the shock front based on measurements of the shock wave thickness
$\delta _F$.

\section{Method of analysis}

This analysis involved coronal images obtained by LASCO C2 and C3
onboard the SOHO spacecraft \citep{Brueckner1995}, presented as
difference brightness $\Delta P = P(t) - P(t_0)$, where $P(t_0)$
is undisturbed brightness at a moment $t_0$, before the event
considered; $P(t)$ is disturbed brightness at $t > t_0$. We used
calibrated LASCO images with the total brightness $P(t)$ expressed
in units of the mean solar brightness (P$_{msb}$).

The difference brightness images were employed to study the
dynamics of the CME and its disturbed region. For this purpose, we
used maps of $\Delta P$ isolines as well as sections along the
Sun's radius at fixed position angles $PA$ and non-radial sections
at various times. In the images, the position angle $PA$ was
counterclockwise from the Sun's northern pole.

At 1.2 R$_\odot < R < 2$ R$_\odot$, we employed polarization
brightness images from the ground-based coronagraph-polarimeter
Mark 4 (Mauna Loa Solar Observatory). As was the case with LASCO
data, these images were expressed in terms of difference
brightness.

\section{Identification of the shock front in front of a CME}

To identify a shock front in CME images is best done by tracing
the process of its formation. Let us examine such a process on the
example of CME 1, which occurred at the W limb on 20 September
1997 at about 10:00 UT. In that event, the coronal ejection had a
distinct three-part structure consisting of a frontal structure
(FS), cavity, and bright core. Figure 1 (three top panels)
presents the difference brightness in the form of isolines for
three subsequent instants of time corresponding to the CME motion.
The shape of the CME frontal
structure is close to a circle (dotted circle in Figure 1).

The process of shock wave formation can be seen in detail on
difference brightness distributions $\Delta P(t,R) = P(t,R) -
P(t_0, R)$ plotted along the direction of the CME motion (dashed
line in the top panels of Figure 1).
Each difference brightness distribution $\Delta P(t,R)$ plotted at
a given instant was (see bottom panel in Figure 1):
\begin{enumerate}
\item normalized to a corresponding maximum value of difference
brightness measured in the vicinity of the frontal structure;
\item displaced along the distance axis $r$ in such a way that
the frontal structure position on the axis coincided with
its position at the moment of its first registration by LASCO C2.
Thus the coordinate system was tied with the frontal structure.
\end{enumerate}

The disturbed region is almost absent at the initial instant
(solid circles) in front of the frontal structure (slanting
hatching in the figure).
At the next moment (empty circles) a
compressed plasma region emerges in front of the FS, bounded by
the shock in its frontal part (crosshatching). The shock moves
faster than the frontal structure and at subsequent moments of
time (diamonds and triangles) is seen to be far ahead of the
frontal structure. Conversely, the anterior boundary of the core
lags behind the frontal structure, because of a lower velocity.

\section{Current sheet and how it is different from the shock front}

Any density inhomogeneity in the magnetized coronal plasma can
only be stationary thanks to magnetic field inhomogeneity, which
is equivalent to the presence of current on the same scale. The
thickness of quiescent current sheet $\delta _I$ expanding due to
diffusion can be estimated, via time $t$, from the following
relation \citep{Dinklage2005}:
\[ \delta _I\approx \rho _e \sqrt{t/\tau _{ep}} \]
where $\tau _{ep}\approx 10^{-2}T^{3/2}/N$ is the mean time
between electron-proton collisions (here $T$ in degrees and $N$ in
cm$^{-3}$).
Let us estimate
$\delta _I$ for our conditions. Assuming, in accordance with
\cite{Mann1999}, that, at $R\approx$ 2.1 R$_\odot$, plasma
temperature $T\approx 1.4\times 10^6$ K, magnetic field $B
\approx$ 0.55 G, and density $N \approx 5\times 10^6$ cm$^{-3}$ we
obtain $\rho _e \sim 10^{-9}$ R$_\odot$ and $\tau _{ep} \approx$ 3
s.
Hence during CME propagation in the corona (for several hours),
the current sheet thickness does not exceed $\sim 100\rho _e$
($\sim 10^{-7}$ R$_\odot$) which is much less than the spatial
resolution of Mark 4 and C2 ($\sim$ 0.02 R$_\odot$). This means
that the minimal measured current sheet thickness is, at best,
close to the spatial resolution of the instruments in use.

However, the motion of the whole current sheet generates in front
a disturbed region due to piled-up background plasma particles and
to excitation of plasma density and magnetic field variations. In
this case, the image brightness jump corresponding to the current
sheet will have an enhanced size due to the effect of the
disturbed region.

The brightness jump in the shock front is also related to the
density inhomogeneity on the scale of the front. Inherently,
however, it differs significantly from the current sheet:
deceleration and heating of the supersonic plasma stream occur in
the shock front.
A disturbed region is absent ahead of the shock front as it
moves at supersonic speed relative to the environment, and hence
the shock front profile does not undergo distortions.


We determined the current sheet thickness at the frontal structure
boundary as double the size of the brightness jump at half the
jump height ($\delta _I$ in the bottom panel in Figure~1).
Besides, it is possible to determine the current sheet thickness
$\delta _{IC}$ at the core boundary.
This value is shown in
Figure~1 only schematically, because the maximum core brightness,
which, in reality, is much larger, was intentionally limited in
the plot. Similarly, Figure 1 illustrates the determination of the
thickness of a shock front the brightness jump in which has a
typical size $\delta _F$.

To understand how we can distinguish the current sheet from a
shock wave, let us consider CME 2 that occurred at the W limb on 2
June 1998 at about 08:00 UT. In that event (in contrast to the
event of 20 September 1997), the CME velocity was lower than the
critical one, and no shock wave formation was observed, at least
in the C2 field of view (i.e. up to $\approx$ 6 R$_\odot$). The
top panel in Figure~2 shows the difference brightness at 10:05 UT
for the event. A three-part structure (FS, cavity and core) may
also be discerned in the event, but there was a rather extended
disturbed region in front of the CME in the direction of its
motion.

Approximating the shape of the frontal structure as a circle (dots
in the top panel of Figure 2) enables us to determine how the
current sheet size changes at the front boundary of the frontal
structure in various directions. For this purpose, we plotted
difference brightness sections from the frontal structure center.
These sections were used to estimate the current sheet size
$\delta _I$. The position of each of these sections was specified
by the angle $\alpha$ drawn from the frontal structure center.
This angle was measured from the direction of CME propagation. The
change in this angle is positive counterclockwise.

In the bottom panel of Figure~2, black circles indicate the
dependence of $\delta _I$ on angle $\alpha$ for CME 2 at 10:05 UT.
There was a developed disturbed region ahead of the frontal
structure (the maximum distance of the frontal structure was
$\approx$ 3.5 R$_\odot$). As a result, $\delta _I$ was nearly five
times as large in the direction of the CME motion than in lateral
directions ($\alpha\approx \pm 100^\circ$) (Figure 2, solid
circles). In CME 1 at 10:19 UT (the maximum distance of the
frontal structure from the solar center at this instant was
$\approx$ 2.2 R$_\odot$), the shock front was not observed yet,
and $\delta _I$ could also be determined at various angles (empty
circles in the bottom panel in Figure~2). The $\delta _I$
thickness is seen to be approximately constant (0.15 R$_\odot$),
only increasing about twofold in the direction of CME motion
($\alpha\approx 0^\circ$).

Hence the development of a disturbed region may result in
increased apparent thickness $\delta _I$. The disturbed region is
only slightly visible and $\delta _I$ is the smallest at large
angles $\alpha$ (in lateral directions).

A similar angle $\alpha$ dependence can also be plotted for the
shock front thickness $\delta _F$. In the bottom panel of Figure
2, crosses mark the $\delta _F(\alpha)$ plot for CME 1 at 11:09
UT, when the maximum distance of the shock front was $\approx$ 4.8
R$_\odot$. Obviously, the behavior of $\delta _F(\alpha)$ differs
from that of $\delta _I(\alpha)$. This difference may be due to
the lack of a disturbed region ahead of the shock front. In that
case, the shock front thickness may depend on local parameters of
the ambient plasma as well as on the velocity component along the
normal to the shock front (which decreases with increasing angle
$\alpha$) -- the shock wave type may change in that case.

\section{Estimating the resolution}

What is the minimum thickness to be recorded by Mark 4 and LASCO
C2 for the current sheet in the corona?

Evidently the effect of widening in the optically thin corona will
be smallest for the current sheet whose size is smallest along the
line of sight. An example is the boundary of the erupting filament
in the form of a thin loop, whose size is sufficiently small along
the line of sight.

The top panel in Figure~3 presents the difference polarization
brightness from Mark 4 data for the CME that commenced at the W
limb on 28 June 2000 at about 19:00 UT. In that event one could
observe a filament eruption easily discernible both in Mark 4 and
LASCO C2 images.
This event allows us to estimate
and compare the Mark 4 and C2 resolutions. For the purpose, we
plotted difference brightness profiles across the filament loop
along the dashed line passing through the filament center
(Figure~3, top panel).
Difference brightness profiles 
normalized to the maximum brightness and shifted in such a manner
that their maxima coincided (bottom panel in Figure~3). The plot
shows that the spatial size of the brightness jump at the filament
boundary remains constant, $\delta _{IC} \approx$ 0.045 R$_\odot$,
over the entire range $R\approx$ 1.3 R$_\odot$ to $R\approx$ 6.1
R$_\odot$. This value is close to the spatial resolution of Mark 4
and C2 ($\sim$ 0.02 R$_\odot$).

%

The main contribution to the jump widening appears to be by the
disturbed region. Thus it is possible to assume that the
observable shock wave thickness ($\delta _F\sim$ 0.2 R$_\odot$ in
the bottom panel of Figure~2) reflects its real size, as there are
no disturbances ahead of the shock front, while the measured shock
wave thickness is essentially larger than the spatial resolution
of the instrument.

Notably, calculations for a simple geometric model of
quasi-spherical shock \citep{Eselevich2008} show that the
observable brightness profile width $\delta _F$ was close to width
$\delta _N$ for the density jump in the shock wave.

\section{Discussion of the energy dissipation mechanism in the shock
wave}

The measured front size $\delta _F$ allows us to answer the
question of whether the mechanism of energy dissipation in the
shock wave is collisionless \citep{Sagdeev1964} or collisional
\citep{Zel'dovich1966}.

If it is collisionless, energy dissipation in a shock wave is
conditioned by collective processes in plasma due to developing
instabilities. In this case, it is quasi-parallel shocks that have
maximum front thickness in the magnetized plasma, that does not
exceed
\[ \delta ^\ast \approx (10-100)\rho_p \]
where $\rho _p$ is the proton Larmor radius calculated from
undisturbed magnetic field directly ahead of the front
\citep{Eselevich1983}.

To estimate the Larmor radius, we take the coronal magnetic field
$B \approx$ 0.5 G and the proton velocity $V = 3\times 10^3$ km/s
for the fastest CME case.
We have: $\delta
^\ast \approx 100\rho _p \sim 10^{-4}$ R$_\odot \ll \delta _F$.
Obviously, the collisionless shock wave front thickness is well
below the resolution limit of modern coronagraphs.
Measurements show, however, that the front thickness
far exceeds this value. This suggests that the dissipation
mechanism in the shock wave is collisional in the corona. In this
case, the shock wave energy dissipation is on the scale of order
of the proton mean free path $\lambda _p$ and hence it is also the
scale that determines the shock wave thickness
\citep{Zel'dovich1966}.

The proton mean free path expressed in solar radii is
\citep{Dinklage2005}:
\begin{equation}
\lambda _p \approx 10^{-7} T^2 / N
\end{equation}
where $T$ and $N$ are respectively the proton
temperature (degrees) and density (cm$^{-3}$) in the undisturbed
plasma immediately ahead of the shock front.

Compare the observable shock wave front thickness $\delta _F$ with
$\lambda _p$ in the corona.
The upper dashed curve in Figure~4 is for $\lambda _p$ calculated
for the proton temperature and density measured by
\cite{Strachan2002}. The lower dashed line is for $\lambda _p$
calculated for the same density and the temperature half as high
as the measurements ($\sim 10^6$ K at 2 R$_\odot$), but suffering
the same decay with distance. These two dashed curves define
roughly the lower and upper boundaries in estimating the free path
depending on the chosen temperature.

The experimental dependence $\delta _F(R)$ was plotted from the
shock wave front thickness measured at different distances for
eight CMEs with velocities $u > u_C$ (symbols in Figure 4).
The mean curve fitting these data is the thin solid line in
Figure~4.
The characteristic
front size is comparable with the free path ($\delta _F \sim
\lambda _p$), at least up to $\sim$5 R$_\odot$. This confirms the
assumption that the dissipation mechanism in the shock wave may be
collisional at these distances. The condition $\delta _F \sim
\lambda _p$ is no valid at greater distances and the shock wave
must become collisionless. We did observe a gradual transition to
formation of a collisionless shock front with thickness $\delta
_F^\ast \ll \lambda _p$ at $R \geq$ 10 R$_\odot$. This will be
dealt with in more detail in a future paper.

Thus, we appear to encounter a rare situation where we can resolve
and examine the collisional shock front structure in plasma.

\section{Conclusions}

The measured thickness of the shock wave, excited ahead of a CME,
far exceeds the spatial resolution of Mark 4 and LASCO C2
coronagraphs. Up to the distance $R\sim$ 5 R$_\odot$ from the
center of the Sun the thickness is of order of the proton mean
free path. This means that the energy dissipation mechanism in a
shock front is collisional.

\begin{acknowledgments}

The work was supported by program No. 16 part 3 of the Presidium
of the Russian Academy of Sciences, program of state support for
leading scientific schools NS-2258.2008.2, and the Russian
Foundation for Basic Research (Project No. 09-02-00165a). The
SOHO/LASCO data used here are produced by a consortium of the
Naval Research Laboratory (USA), Max-Planck-Institut fuer
Aeronomie (Germany), Laboratoire d'Astronomie (France), and the
University of Birmingham (UK). SOHO is a project of international
cooperation between ESA and NASA. The Mark 4 data are courtesy of
the High Altitude Observatory/NCAR.

\end{acknowledgments}

\end{article}

\begin{figure}
 \noindent \center \includegraphics[width=35pc]{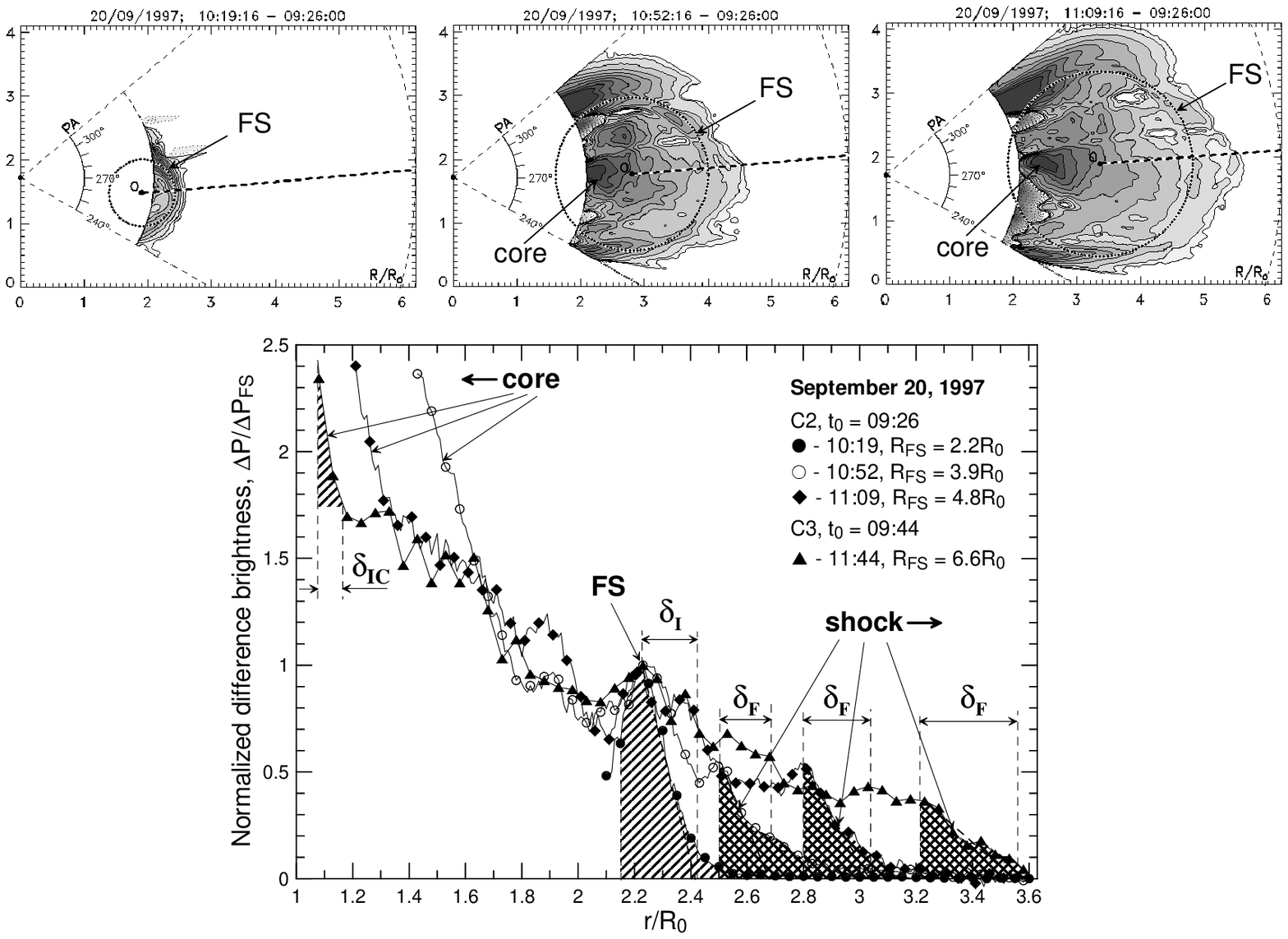}
\caption{CME 1, 20 September 1997, from LASCO C2 and C3 data. Top
panels present difference brightness images for three instants of
time. The bottom panel shows difference brightness distributions
at successive instants of time starting from the frontal structure
center in the directions indicated by the dashed line in the top
panels. These distributions are plotted in the coordinate system
of the frontal structure and normalized to the frontal structure
brightness at the very first instant of time.}
\end{figure}

\begin{figure}
 \noindent \center \includegraphics[width=25pc]{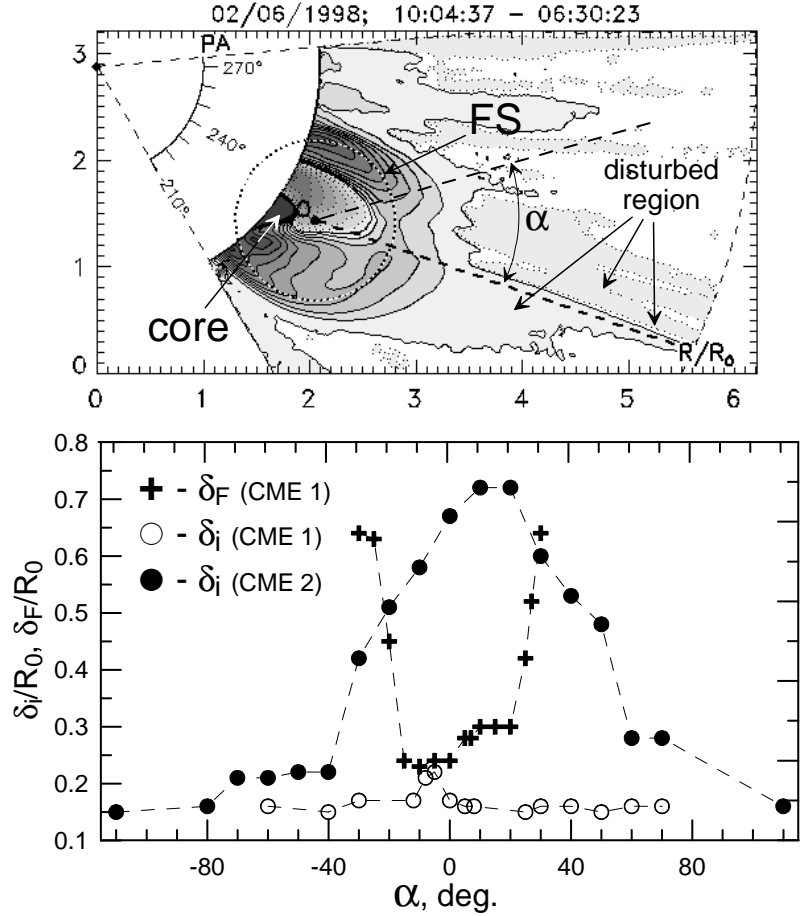}
\caption{The top panel presents difference brightness images for
CME 2 at 10:05 UT on 2 June 1998. The bottom panel: variation in
the observable size of the brightness jump depending on the angle
$\alpha$ measured from the direction of CME propagation at the CME
1 shock front boundary at 11:44 UT (crosses), CME 1 frontal
structure boundary at 10:19 UT (empty circles), CME 2 frontal
structure boundary at 10:05 UT (solid circles).}
\end{figure}

\begin{figure}
 \noindent \center \includegraphics[width=25pc]{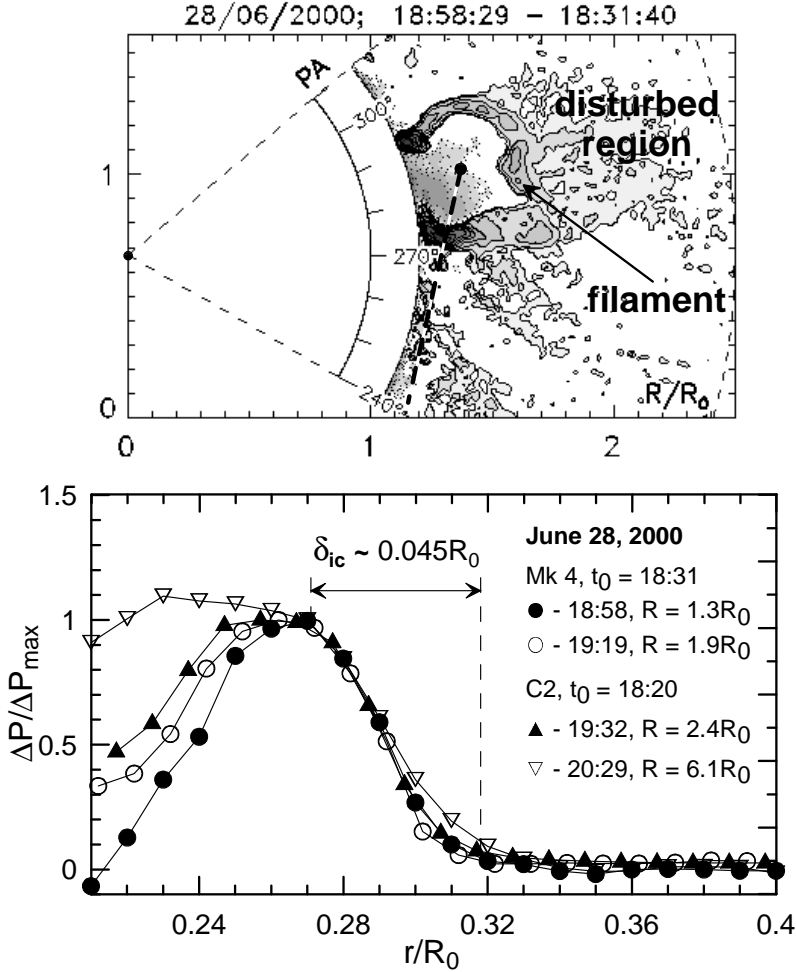}
\caption{CME with filament eruption on 28 June 2000. Top panel:
difference brightness image at 18:59 UT (Mark 4 data). Bottom
panel: difference brightness distributions across the filament in
the lateral direction at various moments during the eruption (Mark
4 and LASCO C2 data).}
\end{figure}

\begin{figure}
 \noindent \center \includegraphics[width=25pc]{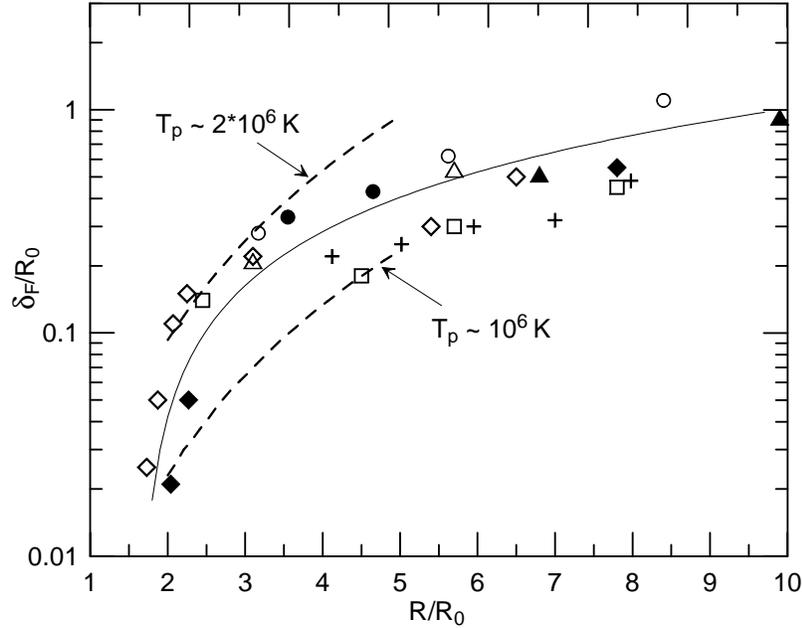}
\caption{The CME-generated shock wave thickness $\delta _F$
variation with distance $R$ from the solar center, in eight CMEs
with high velocities: empty squares -- 20 September 1997, $PA$ =
20$^\circ$; solid circles -- 11 June 1998, $PA$ = 80$^\circ$;
crosses -- 3 March 2000, PA = 230$^\circ$; solid triangles -- 28
June 2000, $PA$ = 270$^\circ$; solid diamonds -- 22 November 2001,
$PA$ = 247-254$^\circ$; empty triangles -- 21 April 2002, $PA$ =
270$^\circ$; empty diamonds -- 26 October 2003, $PA$ =
265-290$^\circ$; empty circles -- 4 November 2003, $PA$ =
238$^\circ$ (from Mark 4 and LASCO C2, C3 data). The dashed curves
are the proton mean free path $\lambda _p$ calculated for two
proton temperatures.
}
\end{figure}

\end{document}